\documentclass[aps,
prl,
twocolumn,
superscriptaddress,
showpacs, 
amsmath,
amssymb]{revtex4-1}

\newcommand*{\btag}{\ensuremath{{B}_{\text{tag}} }}
\newcommand*{\bsig}{\ensuremath{{B}_{\text{sig}} }}

\newcommand*{\R}{\ensuremath{\mathcal{R} }}
\newcommand*{\RD}{\ensuremath{\mathcal{R}(D) }}
\newcommand*{\RDSt}{\ensuremath{\mathcal{R}(D^{\ast}) }}
\newcommand*{\RDall}{\ensuremath{\mathcal{R}(D^{(\ast)}) }}

\newcommand*{\eecl}{\ensuremath{E_{\text{ECL}} }}
\newcommand*{\costheta}{\ensuremath{\cos\theta_{B,D^{(\ast)}\ell} }}

\newcommand*{\tauDec}{\ensuremath{\tau^- \to \ell^- \bar{\nu}_{\ell} \nu_{\tau} }}

\newcommand*{\YFS}{\ensuremath{\PgUc }}

\usepackage{graphicx}
\usepackage{dcolumn}
\usepackage{bm}
\usepackage{hyperref}
\usepackage[mathlines]{lineno}
\usepackage{subfigure}
\usepackage{multirow}
\usepackage{color}
\usepackage{url}
\usepackage{xcolor}
\usepackage[italic]{hepnames}

\hypersetup{
}

\graphicspath{{imports/figures/}}

\begin{document}

\preprint{\vbox{ 
			      \hbox{Version 4.2}
			      \hbox{Belle Preprint 2019-18}
			      \hbox{KEK Preprint 2019-40}
                  \hbox{Intended for {\it PRL}}
}}

\title{Measurement of \RD\ and \RDSt\ with a semileptonic tagging method}


\date{\today}

\noaffiliation
\affiliation{University of the Basque Country UPV/EHU, 48080 Bilbao}
\affiliation{Beihang University, Beijing 100191}
\affiliation{Brookhaven National Laboratory, Upton, New York 11973}
\affiliation{Budker Institute of Nuclear Physics SB RAS, Novosibirsk 630090}
\affiliation{Faculty of Mathematics and Physics, Charles University, 121 16 Prague}
\affiliation{Chonnam National University, Gwangju 61186}
\affiliation{University of Cincinnati, Cincinnati, Ohio 45221}
\affiliation{Deutsches Elektronen--Synchrotron, 22607 Hamburg}
\affiliation{Duke University, Durham, North Carolina 27708}
\affiliation{Key Laboratory of Nuclear Physics and Ion-beam Application (MOE) and Institute of Modern Physics, Fudan University, Shanghai 200443}
\affiliation{Justus-Liebig-Universit\"at Gie\ss{}en, 35392 Gie\ss{}en}
\affiliation{II. Physikalisches Institut, Georg-August-Universit\"at G\"ottingen, 37073 G\"ottingen}
\affiliation{SOKENDAI (The Graduate University for Advanced Studies), Hayama 240-0193}
\affiliation{Department of Physics and Institute of Natural Sciences, Hanyang University, Seoul 04763}
\affiliation{University of Hawaii, Honolulu, Hawaii 96822}
\affiliation{High Energy Accelerator Research Organization (KEK), Tsukuba 305-0801}
\affiliation{J-PARC Branch, KEK Theory Center, High Energy Accelerator Research Organization (KEK), Tsukuba 305-0801}
\affiliation{Forschungszentrum J\"{u}lich, 52425 J\"{u}lich}
\affiliation{IKERBASQUE, Basque Foundation for Science, 48013 Bilbao}
\affiliation{Indian Institute of Science Education and Research Mohali, SAS Nagar, 140306}
\affiliation{Indian Institute of Technology Bhubaneswar, Satya Nagar 751007}
\affiliation{Indian Institute of Technology Guwahati, Assam 781039}
\affiliation{Indian Institute of Technology Hyderabad, Telangana 502285}
\affiliation{Indian Institute of Technology Madras, Chennai 600036}
\affiliation{Indiana University, Bloomington, Indiana 47408}
\affiliation{Institute of High Energy Physics, Chinese Academy of Sciences, Beijing 100049}
\affiliation{Institute of High Energy Physics, Vienna 1050}
\affiliation{Institute for High Energy Physics, Protvino 142281}
\affiliation{INFN - Sezione di Napoli, 80126 Napoli}
\affiliation{INFN - Sezione di Torino, 10125 Torino}
\affiliation{Advanced Science Research Center, Japan Atomic Energy Agency, Naka 319-1195}
\affiliation{J. Stefan Institute, 1000 Ljubljana}
\affiliation{Institut f\"ur Experimentelle Teilchenphysik, Karlsruher Institut f\"ur Technologie, 76131 Karlsruhe}
\affiliation{Kavli Institute for the Physics and Mathematics of the Universe (WPI), University of Tokyo, Kashiwa 277-8583}
\affiliation{Kennesaw State University, Kennesaw, Georgia 30144}
\affiliation{King Abdulaziz City for Science and Technology, Riyadh 11442}
\affiliation{Department of Physics, Faculty of Science, King Abdulaziz University, Jeddah 21589}
\affiliation{Kitasato University, Sagamihara 252-0373}
\affiliation{Korea Institute of Science and Technology Information, Daejeon 34141}
\affiliation{Korea University, Seoul 02841}
\affiliation{Kyungpook National University, Daegu 41566}
\affiliation{LAL, Univ. Paris-Sud, CNRS/IN2P3, Universit\'{e} Paris-Saclay, Orsay 91898}
\affiliation{\'Ecole Polytechnique F\'ed\'erale de Lausanne (EPFL), Lausanne 1015}
\affiliation{P.N. Lebedev Physical Institute of the Russian Academy of Sciences, Moscow 119991}
\affiliation{Ludwig Maximilians University, 80539 Munich}
\affiliation{Luther College, Decorah, Iowa 52101}
\affiliation{University of Maribor, 2000 Maribor}
\affiliation{Max-Planck-Institut f\"ur Physik, 80805 M\"unchen}
\affiliation{School of Physics, University of Melbourne, Victoria 3010}
\affiliation{University of Mississippi, University, Mississippi 38677}
\affiliation{University of Miyazaki, Miyazaki 889-2192}
\affiliation{Moscow Physical Engineering Institute, Moscow 115409}
\affiliation{Moscow Institute of Physics and Technology, Moscow Region 141700}
\affiliation{Graduate School of Science, Nagoya University, Nagoya 464-8602}
\affiliation{Kobayashi-Maskawa Institute, Nagoya University, Nagoya 464-8602}
\affiliation{Universit\`{a} di Napoli Federico II, 80055 Napoli}
\affiliation{Nara Women's University, Nara 630-8506}
\affiliation{National Central University, Chung-li 32054}
\affiliation{National United University, Miao Li 36003}
\affiliation{Department of Physics, National Taiwan University, Taipei 10617}
\affiliation{H. Niewodniczanski Institute of Nuclear Physics, Krakow 31-342}
\affiliation{Nippon Dental University, Niigata 951-8580}
\affiliation{Niigata University, Niigata 950-2181}
\affiliation{Novosibirsk State University, Novosibirsk 630090}
\affiliation{Osaka City University, Osaka 558-8585}
\affiliation{Pacific Northwest National Laboratory, Richland, Washington 99352}
\affiliation{Panjab University, Chandigarh 160014}
\affiliation{Peking University, Beijing 100871}
\affiliation{University of Pittsburgh, Pittsburgh, Pennsylvania 15260}
\affiliation{Punjab Agricultural University, Ludhiana 141004}
\affiliation{Theoretical Research Division, Nishina Center, RIKEN, Saitama 351-0198}
\affiliation{University of Science and Technology of China, Hefei 230026}
\affiliation{Seoul National University, Seoul 08826}
\affiliation{Showa Pharmaceutical University, Tokyo 194-8543}
\affiliation{Soongsil University, Seoul 06978}
\affiliation{University of South Carolina, Columbia, South Carolina 29208}
\affiliation{Sungkyunkwan University, Suwon 16419}
\affiliation{School of Physics, University of Sydney, New South Wales 2006}
\affiliation{Department of Physics, Faculty of Science, University of Tabuk, Tabuk 71451}
\affiliation{Tata Institute of Fundamental Research, Mumbai 400005}
\affiliation{Department of Physics, Technische Universit\"at M\"unchen, 85748 Garching}
\affiliation{Department of Physics, Tohoku University, Sendai 980-8578}
\affiliation{Earthquake Research Institute, University of Tokyo, Tokyo 113-0032}
\affiliation{Department of Physics, University of Tokyo, Tokyo 113-0033}
\affiliation{Tokyo Institute of Technology, Tokyo 152-8550}
\affiliation{Tokyo Metropolitan University, Tokyo 192-0397}
\affiliation{Virginia Polytechnic Institute and State University, Blacksburg, Virginia 24061}
\affiliation{Wayne State University, Detroit, Michigan 48202}
\affiliation{Yamagata University, Yamagata 990-8560}
\affiliation{Yonsei University, Seoul 03722}
  \author{G.~Caria}\affiliation{School of Physics, University of Melbourne, Victoria 3010} 
   \author{P.~Urquijo}\affiliation{School of Physics, University of Melbourne, Victoria 3010} 
  \author{I.~Adachi}\affiliation{High Energy Accelerator Research Organization (KEK), Tsukuba 305-0801}\affiliation{SOKENDAI (The Graduate University for Advanced Studies), Hayama 240-0193} 
  \author{H.~Aihara}\affiliation{Department of Physics, University of Tokyo, Tokyo 113-0033} 
  \author{S.~Al~Said}\affiliation{Department of Physics, Faculty of Science, University of Tabuk, Tabuk 71451}\affiliation{Department of Physics, Faculty of Science, King Abdulaziz University, Jeddah 21589} 
  \author{D.~M.~Asner}\affiliation{Brookhaven National Laboratory, Upton, New York 11973} 
  \author{H.~Atmacan}\affiliation{University of South Carolina, Columbia, South Carolina 29208} 
  \author{T.~Aushev}\affiliation{Moscow Institute of Physics and Technology, Moscow Region 141700} 
  \author{V.~Babu}\affiliation{Deutsches Elektronen--Synchrotron, 22607 Hamburg} 
  \author{I.~Badhrees}\affiliation{Department of Physics, Faculty of Science, University of Tabuk, Tabuk 71451}\affiliation{King Abdulaziz City for Science and Technology, Riyadh 11442} 
% ...
\collaboration{The Belle Collaboration}


\begin{abstract}
The experimental results on the ratios of branching fractions $\RD = {\cal B}(\bar{B} \to D \tau^- \bar{\nu}_{\tau})/{\cal B}(\bar{B} \to D \ell^- \bar{\nu}_{\ell})$ and $\RDSt = {\cal B}(\bar{B} \to D^* \tau^- \bar{\nu}_{\tau})/{\cal B}(\bar{B} \to D^* \ell^- \bar{\nu}_{\ell})$, where $\ell$ denotes an electron or a muon, show a long-standing discrepancy with the Standard Model predictions, and might hint to a violation of lepton flavor universality. We report a new simultaneous measurement of \RD\ and \RDSt, based on a data sample containing $772 \times 10^6$ $B\bar{B}$ events recorded at the \YFS\ resonance with the Belle detector at the KEKB $e^+ e^-$ collider. In this analysis the tag-side $B$ meson is reconstructed in a semileptonic decay mode and the signal-side $\tau$ is reconstructed in a purely leptonic decay. The measured values are $\RD\ = 0.307 \pm 0.037 \pm 0.016$ and $\RDSt\ = 0.283 \pm 0.018 \pm 0.014$, where the first uncertainties are statistical and the second are systematic. These results are in agreement with the Standard Model predictions within $0.2$, $1.1$ and $0.8$ standard deviations for \RD, \RDSt\ and their combination, respectively. This work constitutes the most precise measurements of \RD\ and \RDSt\ performed to date as well as the first result for \RD\ based on a semileptonic tagging method. 
\end{abstract}


\maketitle


Semitauonic $B$ meson decays, involving the transition $b \to c \tau \nu_{\tau}$, are sensitive probes for physics beyond the Standard Model (SM). Any difference in the branching fraction of these processes with respect to the SM prediction would violate lepton flavor universality, which enforces equal coupling of the gauge bosons to the three lepton generations. Indeed, in many models beyond the SM, new interactions with enhanced coupling to the third family are postulated. Among such new mediators, charged Higgs bosons, which appear in supersymmetry~\cite{Martin:1997ns} and other models with two Higgs doublets~\cite{Gunion:1989we}, may contribute measurably to the $b \to c \tau \nu_{\tau}$ decay rate due to the large masses of the $\tau$ and the $b$ quark. Similarly, leptoquarks~\cite{Buchmuller:1986zs}, which carry both lepton and baryon numbers, may also contribute to this process. 

The ratios of branching fractions,
\begin{eqnarray}
    \RDall = \frac{{\cal B}(\bar{B} \to D^{(*)} \tau^- \bar{\nu}_{\tau})}{{\cal B}(\bar{B} \to D^{(*)} \ell^- \bar{\nu}_{\ell})}
\end{eqnarray}
%
where the denominator represents the average of electron and muon modes,
are typically measured instead of the absolute branching fractions of $\bar{B} \to D^{(*)} \tau^- \bar{\nu}_{\tau}$ to reduce common systematic uncertainties, such as those due to the detection efficiency, the magnitude of the quark-mixing matrix element $|V_{cb}|$, and the semileptonic decay form factors. Hereafter, $\bar{B} \to D^{(*)} \tau^- \bar{\nu}_{\tau}$~\cite{note} and $\bar{B} \to D^{(*)} \ell^- \bar{\nu}_{\ell}$ will be referred to as the signal and normalization modes, respectively. The SM calculations for these ratios, performed by several groups~\cite{Bigi:2016mdz,Bernlochner:2017jka, Bigi:2017jbd, Jaiswal:2017rve}, are averaged by HFLAV~\cite{HFLAV} to obtain
$\RD      = 0.299 \pm 0.003$ and
$\RDSt    = 0.258 \pm 0.005$.

Semitauonic $B$ decays were first observed by Belle in 2007~\cite{PhysRevLett.99.191807}, with subsequent studies reported by Belle~\cite{PhysRevD.82.072005, huschle2015, Sato:2016svk, PhysRevLett.118.211801, *Hirose:2016wfn}, BaBar~\cite{lees2012, *lees2013}, and LHCb~\cite{aaij2015, Aaij:2017deq}.
The average values of the experimental results, excluding the result presented in this Letter, are
$\RD = 0.407 \pm 0.039 \pm 0.024$ and
$\RDSt = 0.306 \pm 0.013 \pm 0.007$~\cite{HFLAV},
where the first uncertainty is statistical and the second is systematic. These values exceed SM predictions by $2.1\sigma$ and $3.0\sigma$, respectively, where $\sigma$ denotes the standard deviation. A combined analysis of \RD\ and \RDSt\, taking correlations into account finds that the deviation from the SM prediction is approximately $3.8\sigma$~\cite{HFLAV}. This large discrepancy must be investigated with complementary and more precise measurements. 

Measurements at the $e^+ e^-$ ``$B$-factory" experiments Belle and BaBar, are commonly performed by first reconstructing one of the $B$ mesons in the $\YFS \to B\bar{B}$ decay, denoted as \btag, using a dedicated tagging algorithm. So far, simultaneous measurements of \RD\ and \RDSt\ at Belle and BaBar have been performed using hadronic tagging methods on both $B^0$ and $B^+$ decays~\cite{huschle2015,lees2012, lees2013}, while only $\R(D^{*+})$ was measured with a semileptonic tagging method~\cite{Sato:2016svk}. In this Letter, we report the first measurement of \RD\ using the semileptonic tagging method, and we update or measurement of \RDSt\ by combining results of $B^0$ and $B^+$ decays with a more efficient tagging algorithm. Our previous measurement of $\R(D^{*+})$ with a semileptonic tagging method is therefore superseded by this work.

We use the full \YFS\ data sample containing $772 \times 10^6$ $B \bar{B}$ events recorded with the Belle detector~\cite{Abashian:2000cg} at the KEKB $e^+ e^-$ collider~\cite{KUROKAWA20031}.
Belle was a general-purpose magnetic spectrometer, which consisted of a silicon vertex detector (SVD), a 50-layer central drift chamber (CDC), an array of aerogel threshold Cherenkov counters (ACC), time-of-flight scintillation counters (TOF), and an electromagnetic calorimeter (ECL) comprising CsI(Tl) crystals. These components were located inside a superconducting solenoid coil that provided a 1.5 T magnetic field. An iron flux-return yoke located outside the coil was instrumented to detect $K_L^0$ mesons and muons (KLM). The detector is described in detail elsewhere~\cite{Abashian:2000cg}.
 
To determine the reconstruction efficiency and probability density functions (PDFs)  for signal, normalization, and background modes, we use Monte Carlo (MC) simulated events generated with the EvtGen event generator~\cite{LANGE2001152}. The detector response is simulated with the GEANT3 package~\cite{Brun:1987ma}. 

Semileptonic $B \to D^{(*)} \ell \nu$ decays are generated with the HQET2 EvtGen package, based on the CLN parametrization~\cite{Caprini1997}. As the measured parameters of the model have been updated since our MC sample was generated, we apply an event-by-event correction factor obtained by taking the ratio of differential decay rates in the updated CLN parameters compared to those used in the MC. For the MC samples of $B \to D^{**} \ell \nu$ decays, we used the ISGW2 EvtGen package, based on the quark model described in Ref.~\cite{Scora1995}. This model has been superseded by the LLSW model~\cite{Leibovich1998}; thus we weight events with a correction factor based on the ratio of the analytic predictions of LLSW and MC distributions generated with ISGW2. Here, $D^{**}$ denotes the orbitally excited states $D_1$, $D_2^*$, $D_1'$, and $D_0^*$.  We consider $D^{**}$ decays to a $D^{(*)}$ and a pion, a $\rho$ or an $\eta$ meson, or a pair of pions, where branching fractions are based on quantum number, phase-space, and isospin arguments. The sizes of the inclusive $\YFS \to B\bar{B}$ MC sample and the dedicated $B \to D^{**} \ell \nu $ MC sample correspond to about 10 times and 5 times the integrated luminosity of the \YFS\ data sample, respectively.


The \btag\ is reconstructed using a hierarchical algorithm based on boosted decision trees (BDT)~\cite{Keck2019} in $D^{} \ell \bar{\nu}_{\ell}$ and $D^{*} \ell \bar{\nu}_{\ell}$ channels, where $\ell = e, \mu$. The BDT classifier assigns to each \btag\ candidate a probability of representing a well-reconstructed $B$ meson. The range of the BDT classifier extends from 0 to 1, with well-reconstructed candidates having the highest values.  We select \btag\ candidates with a BDT classifier output greater than $10^{-1.5}$. We suppress ${B} \to D^{\ast} \tau (\to \ell \nu \nu) \nu  $ events on the \btag\ side by applying a selection on \costheta.
This variable corresponds to the cosine of the angle between the momenta of the $B$ meson and the $D^{(*)} \ell$ system in the \YFS\ rest frame, under the assumption that only one massless particle is not reconstructed:
\begin{eqnarray}
     \costheta \equiv
    \frac
    {2E_{\rm beam} E_{D^{(*)} \ell} - m_B^2 - m_{D^{(*)} \ell}^2}
    {2 |\bm{p}_B| |\bm{p}_{D^{(*)} \ell}|}.
    \label{eq:cos_bdstrl}
\end{eqnarray}
Here $E_{\rm beam}$ is the beam energy, and $E_{D^* \ell}$, $\bm{p}_{D^* \ell}$, and $m_{D^* \ell}$ are the energy, momentum, and mass, of the $D^* \ell$ system, respectively. The quantities $m_B$ and $|\bm{p}_B|$ are the nominal $B$ meson mass~\cite{PDG} and momentum, respectively. All quantities are evaluated in the \YFS\ rest frame. 

Correctly reconstructed $B \to D^{(*)} \ell \nu$ decays are expected to have a value of  \costheta\ between $-1$ and $+1$. Correctly reconstructed as well as misreconstructed $B \to D^{(*)} \tau \nu$ decays generally have \costheta\ values below $-1$ due to the presence of additional missing particles. To account for detector resolution effects we apply the requirement $ -2.0 <  \costheta < 1.0$ for the \btag. 

In each event with a selected \btag\ candidate, we search for the opposite-flavor signature $D^{(*)} \ell$ among the remaining tracks and calorimeter clusters, since we only reconstruct pure leptonic tau decays $\tau \to \ell \bar{\nu} \nu$. We define four disjoint data samples, denoted $D^{+} \ell^{-}$, $D^{0} \ell^{-}$, $D^{*+} \ell^{-}$, and $D^{*0} \ell^{-}$.

Charged particle tracks are reconstructed with the SVD and CDC by requiring a point of closest approach to the interaction point smaller than 5.0 cm along the direction of the $e^+$ beam and 2.0 cm in the direction perpendicular to it. These requirements do not apply to the pions from $K_S^0$ decays. Electrons are identified by a combination of the specific ionization ($dE/dx$) in the CDC, the ratio of the cluster energy in the ECL to the track momentum measured with the CDC, the response of the ACC, the cluster shape in the ECL, and the match between positions of the cluster and the track at the ECL. To recover bremsstrahlung photons from electrons, we add the four-momentum of each photon detected within a cone of 0.05 rad of the original track direction to the electron momentum. Muons are identified by the track penetration depth and hit distribution in the KLM. Charged kaons are identified by combining information from the $dE/dx$ measured in the CDC,  the flight time measured with the TOF, and the response of the ACC. We do not apply any particle identification criteria for charged pion candidates.

Candidate $K_S^0$ mesons are formed by combining two oppositely charged tracks with pion mass hypotheses. We require their invariant mass to lie within $\pm$15 MeV/$c^2$ of the nominal $K^0$ mass~\cite{PDG}, which corresponds to approximately seven times the reconstructed mass resolution. Further selection is performed with an algorithm based on a neural network~\cite{FEINDT2006190}.

Photons are measured as an electromagnetic cluster in the ECL with no associated charged track. Neutral pions are reconstructed in the $\pi^0 \to \gamma \gamma$ channel, and their energy resolution is improved by performing a mass-constrained fit of the two photon candidates to the nominal  $\pi^0$  mass~\cite{PDG}. For neutral pions from $D$ decays, we require the daughter photon energies to be greater than 50 MeV and their asymmetry to be less than 0.6 in the laboratory frame, the cosine of the angle between two photons to be greater than zero, and the $\gamma \gamma$ invariant mass to be within $[-15, +10]$ MeV/$c^2$ of the nominal $\pi^0$ mass, which corresponds to approximately $\pm 1.8$ times the resolution. Low-energy  $\pi^0$  candidates from $D^*$ are reconstructed using less restrictive energy requirements: one photon must have an energy of at least 50 MeV, while the other must have a minimum energy of 20 MeV.  We also require a narrower window around the diphoton invariant mass to compensate for the lower photon-energy requirement: within 10 MeV/$c^2$ of the nominal $\pi^0$ mass, which corresponds to approximately $\pm 1.6$  times the resolution.

Neutral $D$ mesons are reconstructed in the following decay modes:
$D^0 \to K^- \pi^+ \pi^0$,
$K^- \pi^+ \pi^+ \pi^- $,
$K^- \pi^+$,
$K_S^0 \pi^+ \pi^-$,
$K_S^0 \pi^0$,
$K_S^0 K^+ K^-$,
$K^+ K^-$, and
$\pi^+ \pi^-$.
Similarly, charged $D$ mesons are reconstructed in the following modes:
$D^+ \to K^- \pi^+ \pi^+$,
$K_S^0 \pi^+ \pi^0$,
$K_S^0 \pi^+ \pi^+ \pi^-$,
$K_S^0 \pi^+$,
$K^- K^+ \pi^+$, and
$K_S^0 K^+$.
The combined branching fractions for reconstructed channels are 30\% and 22\% for $D^0$ and $D^+$, respectively. For $D$ decays without a $\pi^0$ in the final state,  we require the invariant mass of the reconstructed candidates to be within 15 MeV/$c^2$ of the nominal $D^0$ or $D^+$ mass, which corresponds to a window of approximately $\pm 2.8$ times the resolution. In the case of channels with a $\pi^0$ in the final state, which have worse mass resolution, we require a wider window: from $-45$ to $+30$ MeV/$c^2$ around the nominal $D^0$ mass, and  from $-36$ to $+24$ MeV/$c^2$ around the nominal $D^+$ mass. These windows correspond to approximately [$-1.1, +1.6$] and [$-1.0, +1.4$] times the resolution, respectively. Candidate $D^{*+}$ mesons are reconstructed in the channels $D^0 \pi^+$  and $D^+ \pi^0$, and $D^{*0}$ in the channel $D^0 \pi^0$. We do not consider the $D^{*0} \to D^0 \gamma$ decay channel due to its higher background level.

We require the mass difference $D^*-D$ be within 2.5 MeV/$c^2$ for the $D^{*+} \to D^0 \pi^+$ decay mode, and within 2.0 MeV/$c^2$ for the $D^{*+} \to D^+ \pi^0$ and $D^{*0} \to D^0 \pi^0$ decay modes. These windows correspond to $\pm 3.0$ and $\pm 1.9$ times the resolution, respectively. We require a tighter mass window in the $D^{*}$ modes that contain a low-momentum (``slow") $\pi^0$ to suppress the large background arising from misreconstructed neutral pions.

On the signal side, we require \costheta\ to be less than 1.0 and the $D^{(*)}$ momentum in the \YFS\ rest frame to be less than 2.0 GeV/$c$.
Finally, we require that events contain no extra prompt charged tracks, $K_S^0$ candidates, or $\pi^0$ candidates, which are reconstructed with the same criteria as those used for the $D$ candidates. All selection criteria used for event reconstruction have been the subject of optimization studies. When multiple \btag\ or  \bsig\ candidates are found in an event, we first select the \btag\ candidate with the highest tagging classifier output, and then the \bsig\ candidate with the highest p-value from the vertex fit of the $B$ candidate's charm daughter.


To distinguish signal and normalization events from background processes, we use the sum of the energies of neutral clusters detected in the ECL that are not associated with any reconstructed particles, denoted as \eecl.
To mitigate the varying effects of photons related to beam background in the calculation of \eecl, we only include  clusters with energies greater than 50, 100, and 150 MeV, respectively, from the barrel, forward, and backward ECL regions~\cite{Abashian:2000cg}.
Signal and normalization events peak near zero in $E_{\rm ECL}$, while background events populate a wider range. We require that \eecl\ be less than 1.2 GeV.

To separate reconstructed signal and normalization events, we employ a BDT based on the \verb|XGBoost| package~\cite{xgboost16}. The input variables to the BDT are  \costheta; the approximate missing mass squared $m_{\rm miss}^2 = ( E_{\rm beam} - E_{D^{(*)}} - E_\ell )^2 - (\bm{p}_{D^{(*)}} + \bm{p}_{\ell} )^2$; the visible energy $E_{\rm vis} = \sum_i E_i$, where $(E_i, \bm{p}_i)$ is the four-momentum of particle $i$. We do not apply any selection on the BDT classifier output, denoted as $O_{\text{cls}}$; instead we use it as one of the fitting variables for the extraction of \RDall. Signal events have $O_{\text{cls}}$ values near 1, while normalization events have values near 0.

We extract the yields of signal and normalization modes from a two-dimensional (2D) extended maximum-likelihood fit to the variables $O_{\text{cls}}$ and \eecl. The fit is performed simultaneously to the four $D^{(*)} \ell$ samples, and exploits the isospin constraint $R(D^{(*)0}) = R(D^{(*)+})$. The distribution of each sample is described as the sum of several components: $D^{(*)} \tau \nu$,  $D^{(*)} \ell \nu$, feed-down from $D^* \ell (\tau) \nu$ to $D \ell (\tau) \nu$, $D^{**} \ell(\tau) \nu$, and other backgrounds.  The PDFs of these components are determined from MC simulations as 2D histogram templates.
A large fraction of $B \to D^* \ell \nu$ decays from both $B^0$ and $B^+$ are reconstructed in the $D \ell$ samples (denoted feed-down). We leave these two contributions free in the fit and use their fitted yields to correct the MC estimated feed-down rate of $B \to D^* \tau \nu$ decays. The events of the $D^* \ell$ samples that appear as feed-down are treated as a component of the signal or normalization yields. As the probability of $B \to D \ell(\tau) \nu$ decays contributing to the $D^* \ell$ samples is very small, the relative rates of these contributions are fixed to the MC expected values.

The free parameters in the final fit are the yields of signal, normalization, $B \to D^{**} \ell \nu_{\ell}$, and feed-down from $D^* \ell$ to $D \ell$ components. The yields of other backgrounds are fixed to their MC expected values.
The ratios \RDall\ are given by the formula:
\begin{eqnarray}
    \RDall &=&
    \frac{1}{2{\cal B}(\tau^- \to \ell^- \bar{\nu}_{\ell} \nu_{\tau})}
    \cdot
    \frac{\varepsilon_{\rm norm}}{\varepsilon_{\rm sig}}
    \cdot
    \frac{N_{\rm sig}}{N_{\rm norm}},
    \label{eq:cal_rdstr}
\end{eqnarray}
where $\varepsilon_{\rm sig (norm)}$ and $N_{\rm sig (norm)}$ are the detection efficiency including tagging efficiency and yields of signal (normalization) modes and ${\cal B}(\tau^- \to \ell^- \bar{\nu}_{\ell} \nu_{\tau})$ is the average of the world-average branching fractions for $\ell =e$ and $\ell = \mu$.
\begin{figure*}[tbp]
    \includegraphics[width=\columnwidth]{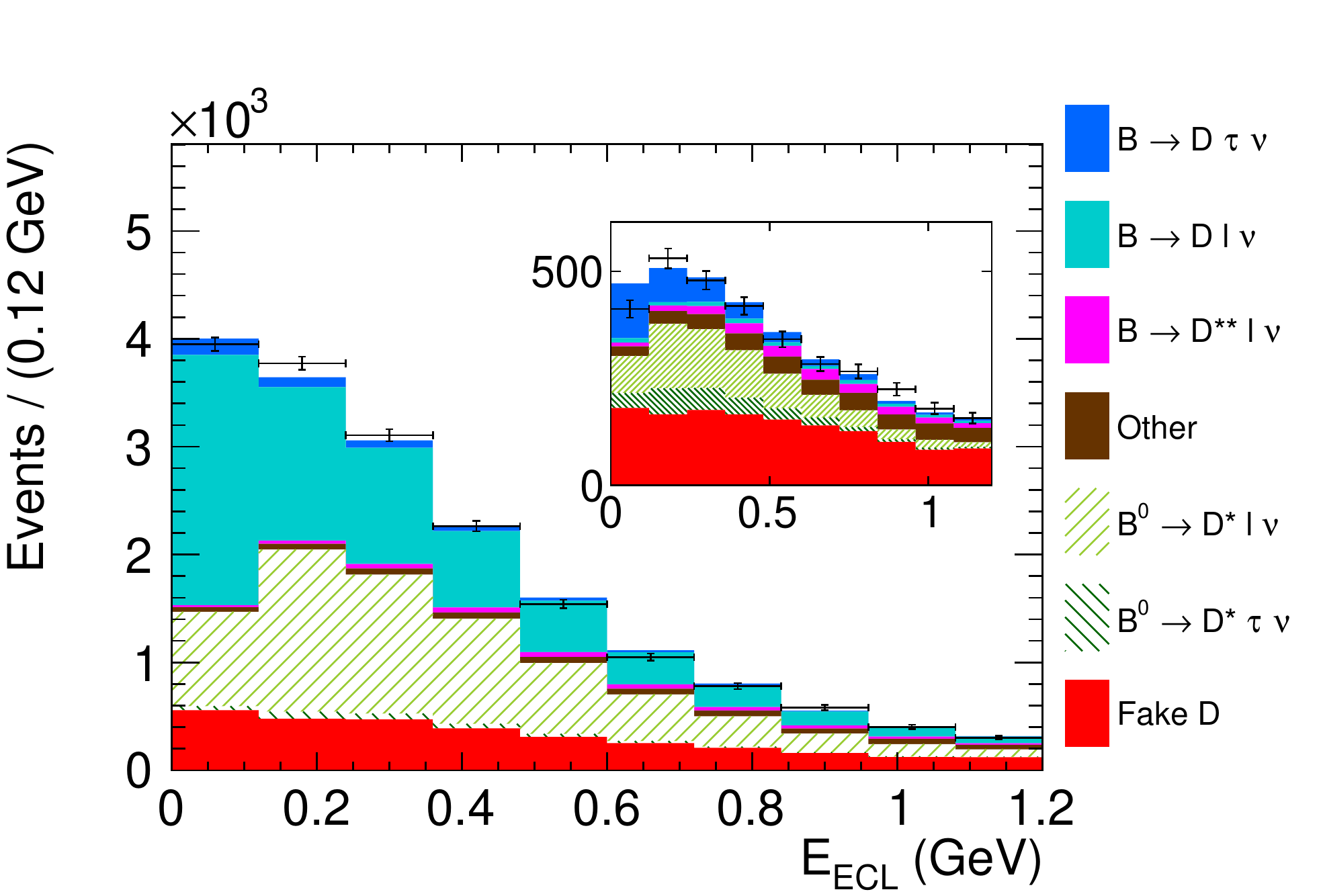}
    \includegraphics[width=\columnwidth]{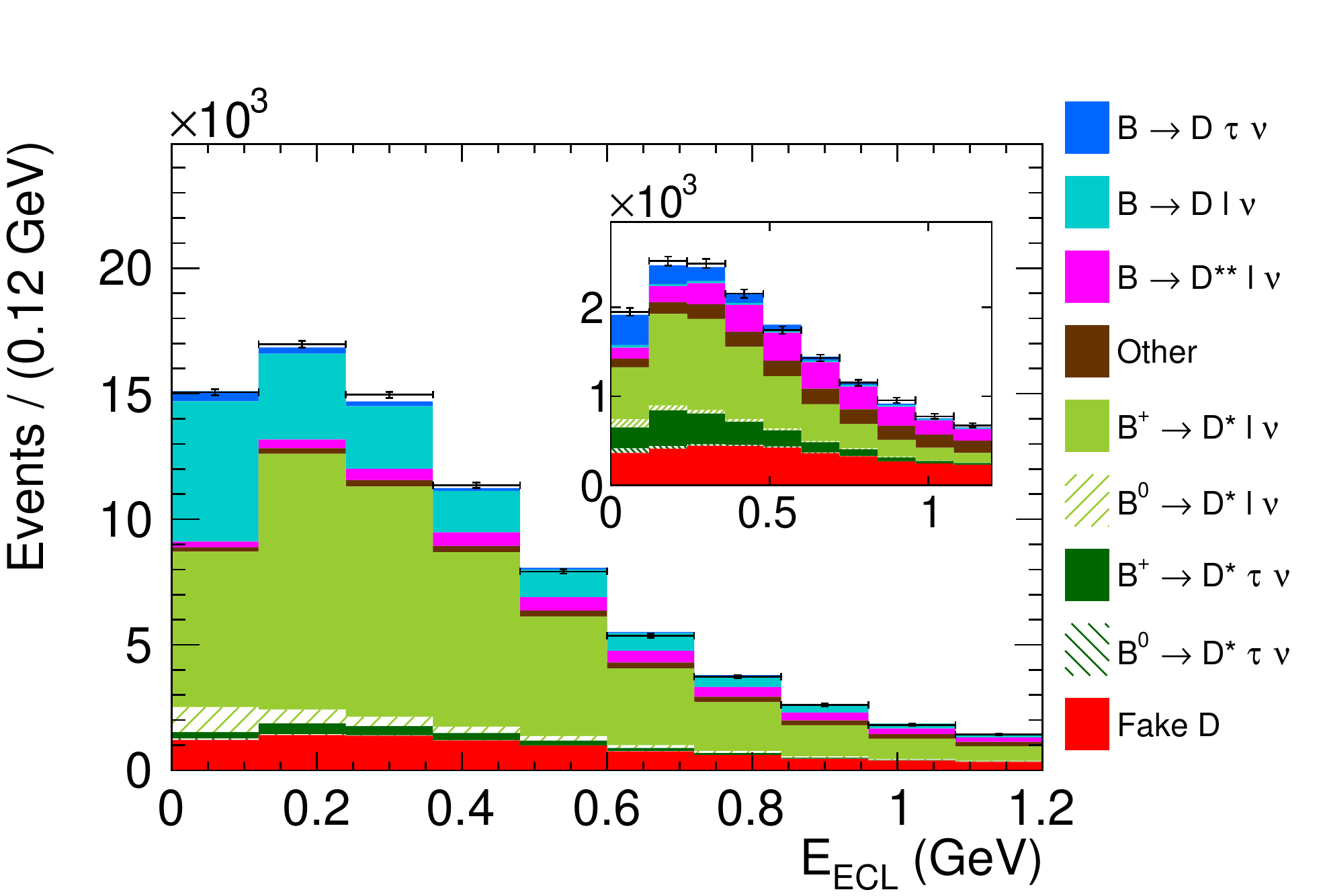}
    \includegraphics[width=\columnwidth]{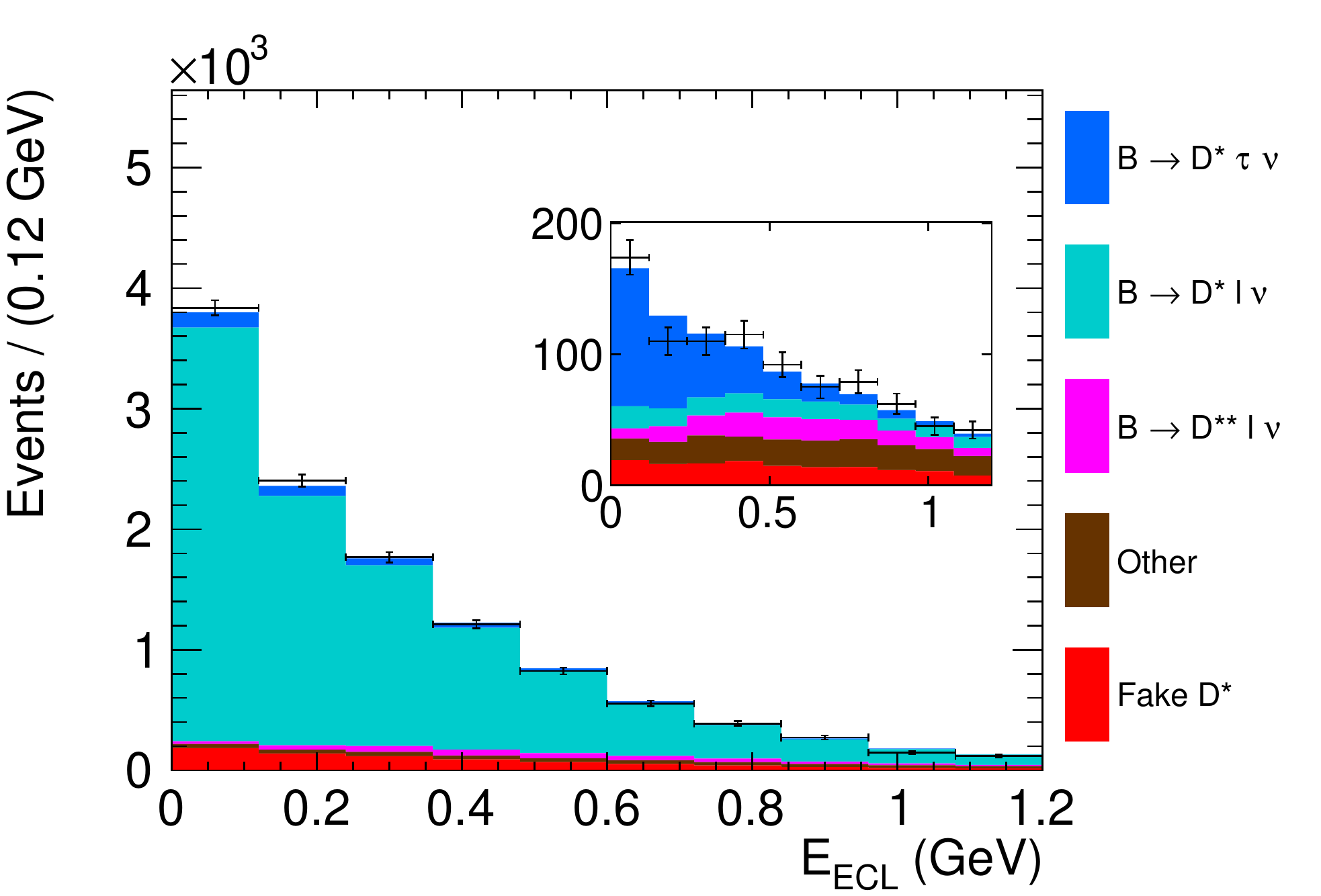}
    \includegraphics[width=\columnwidth]{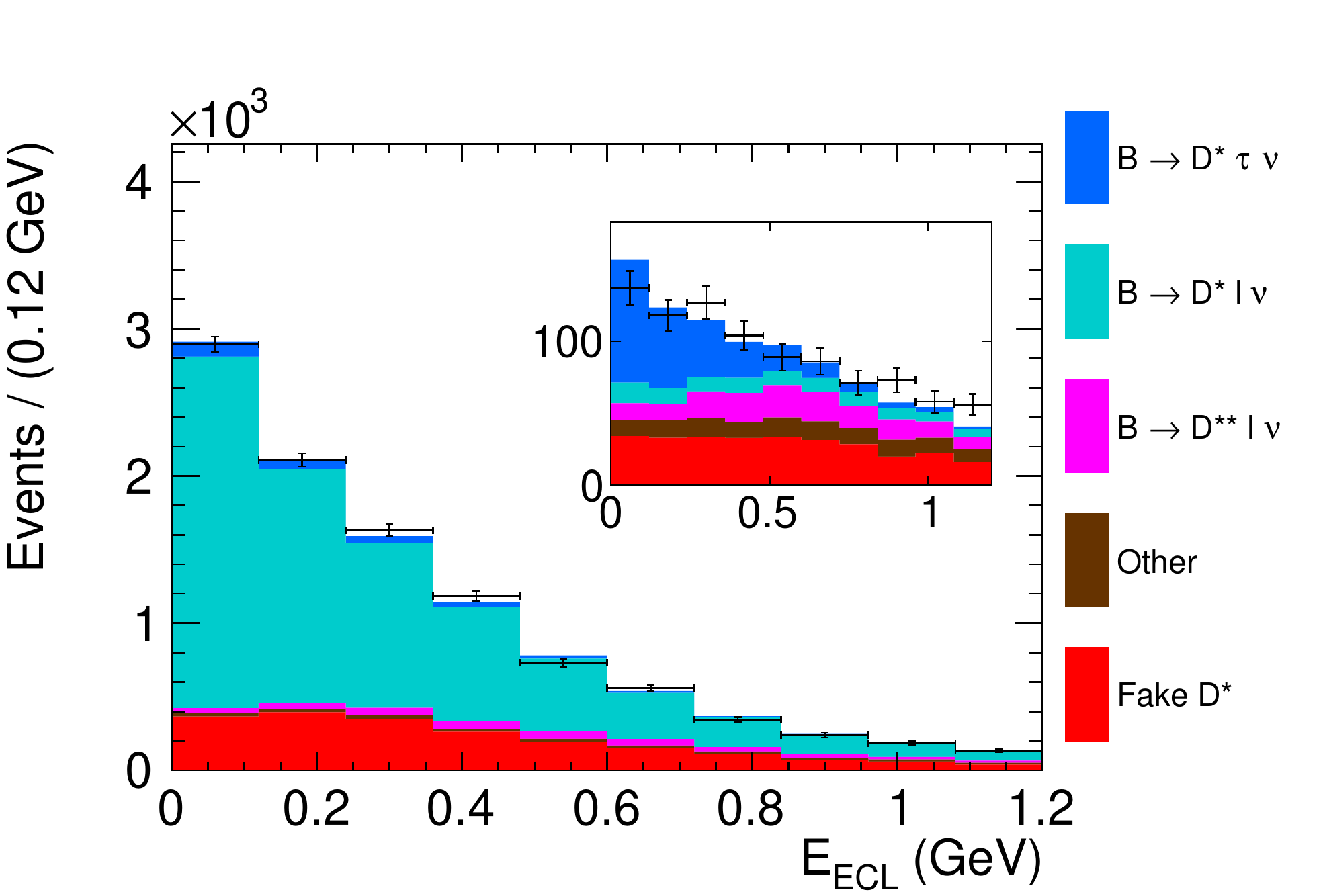}
    \caption{\eecl\ fit projections and data points with statistical uncertainties in the $D^+\ell^-$ (top left), $D^0\ell^-$ (top right), $D^{*+}\ell^-$ (bottom left) and $D^{*0}\ell^-$ (bottom right) samples, for the full classifier region. The signal region, defined by the selection $O_{\text{cls}} >0.9$, is shown in the inset.}
    \label{fig:results_allModes}
\end{figure*}
To improve the accuracy of the MC simulation, we apply a series of correction factors determined from control sample measurements, such as those associated to lepton and hadron identification efficiencies as well as slow pion tracking efficiencies.
Correction factors for the lepton efficiencies are evaluated as a function of the lepton momentum and direction using $e^+ e^- \to e^+ e^- \ell^+ \ell^-$ and $J/\psi \to \ell^+ \ell^-$ decays. 
Furthermore, to determine the expected yield of fake and misreconstructed $D^{(*)}$ mesons, treated as background, we use data sidebands of difference between their nominal and reconstructed mass, and we correct for differences in the reconstruction efficiency of the tagging algorithm between data and MC simulation.

The \eecl\ projections of the fit are shown in Fig.~ \ref{fig:results_allModes}. The fit finds  $\RD = 0.307 \pm 0.037$ and $\RDSt = 0.283 \pm 0.018$, where the error is statistical.


To estimate various systematic uncertainties contributing to \RDall, we vary each fixed parameter 500 times, sampling from a Gaussian distribution built using the value and uncertainty of the parameter. For each variation, we repeat the fit. The associated systematic uncertainty is taken as the standard deviation of the resulting distribution of fitted results. The systematic uncertainties are listed in Table~\ref{tab:sys}. 

In Table~\ref{tab:sys} the label ``$D^{**}$ composition" refers to the uncertainty introduced by the branching fractions of the $B \to D^{**} \ell \nu_{\ell}$ channels and the decays of the $D^{**}$ mesons, which are not well known and hence contribute significantly to the total PDF uncertainty. The uncertainties on the branching fraction of $B \to D^{**} \ell \nu_{\ell}$ are assumed to be $\pm 6\%$ for $D_1$, $\pm 10\%$ for $D_2^*$, $\pm 83\%$ for $D_1'$, and $\pm 100\%$ for $D_0^*$, while the uncertainties on each of the $D^{**}$ decay branching fractions are conservatively assumed to be $\pm 100\%$.

A large systematic uncertainty arises from the limited size of the MC samples. Firstly, this is reflected in the uncertainty of the PDF shapes. To estimate this contribution, we recalculate PDFs for signal, normalization, fake $D^{(*)}$ events, $B \to D^{**} \ell \nu_{\ell}$, feed-down, and other backgrounds by generating toy MC samples from the nominal PDFs according to Poisson statistics, and then repeating the fit with the new PDFs. 
Secondly, the reconstruction efficiency of feed-down events, together with the efficiency ratio of signal to normalization events, are varied within their uncertainties, which are limited by the size of the MC samples as well.

The efficiency factors for the fake $D^{(*)}$ and \btag\ reconstruction are calibrated using collision data. The uncertainties on these factors are affected by the size of the samples used in the calibration. We vary the factors within their errors and extract associated systematic uncertainties.

The effect of the lepton efficiency and fake rate, as well as that due to the slow pion efficiency, do not cancel out in the $\R (D^{(*)})$ ratios. This is due to the different momentum spectra of leptons and charm mesons in the normalization and signal modes. The uncertainties introduced by these factors are included in the total systematic uncertainty.

We include minor systematic contributions from other sources: one related to the parameters that are used for re-weighting the semileptonic $B\to D^{(*)} \ell \nu$ and $B \to D^{**}\ell \nu$ decays; and others from the integrated luminosity, the $B$ production fractions at the $\Upsilon(4S)$, $f^{+-}$ and $f^{00}$, and the branching fractions of $B \to D^{(*)} \ell \nu$, $D$, $D^*$ and  $\tau^- \to \ell^- \bar{\nu}_{\ell} \nu_{\tau}$ decays~\cite{PDG}.
The total systematic uncertainty is estimated by summing the aforementioned contributions in quadrature.

\begin{table}
	\caption{Systematic uncertainties contributing to the \RDall\ results, together with their correlation.}
	\begin{ruledtabular}
	\centering
	\begin{tabular}{p{3cm}ccc}
		Source 								    &  $\Delta\RD\ (\%)$    &  $\Delta\RDSt\ (\%)$ &     Correlation \\
		\hline
		$D^{**}$ composition                    &                     0.76 &                      1.41 &    $-0.41$ \\
		PDF shapes                              &                     4.39 &                      2.25 &    $-0.55$ \\
		Feed-down factors                       &                     1.69 &                      0.44 &    $\quad 0.53$ \\
		Efficiency factors                      &                     1.93 &                      4.12 &    $-0.57$ \\
		Fake $D^{(*)}$ calibration              &                     0.19 &                      0.11 &    $-0.76$ \\
		\btag\ calibration                      &                     0.07 &                      0.05 &    $-0.76$ \\
		Lepton efficiency                       &                     0.36 &                      0.33 &    $-0.83$ \\
		and fake rate                           &                          &                           &             \\
		Slow pion efficiency                    &                     0.08 &                      0.08 &    $-0.98$ \\
		$B$ decay form factors                  &                     0.55 &                      0.28 &    $-0.60$ \\
		Luminosity, $f^{+-}$, $f^{00}$          &                     0.10 &                      0.04 &    $-0.58$ \\
		and $\mathcal{B}(\Upsilon(4S))$         &                          &                           &            \\
		$\mathcal{B}(B \to D^{(*)} \ell \nu)$   &                     0.05 &                      0.02 &    $-0.69$ \\
		$\mathcal{B}(D)$                        &                     0.35 &                      0.13 &    $-0.65$ \\
		$\mathcal{B}(D^*)$                      &                     0.04 &                      0.02 &    $-0.51$ \\
		$\mathcal{B}(\tauDec)$ 				    &                     0.15 &                      0.14 &    $-0.11$ \\
		\hline
		Total                                	&                     5.21 &                      4.94 &    $-0.52$ \\
	\end{tabular}
	\label{tab:sys}
	\end{ruledtabular}
\end{table}


In conclusion, we have measured the ratios $\RDall = {\cal B}(\bar{B} \to D^{(*)} \tau^- \bar{\nu}_{\tau})/{\cal B}(\bar{B} \to D^{(*)} \ell^- \bar{\nu}_{\ell})$, where $\ell$ denotes an electron or a muon, using a semileptonic tagging method and a data sample containing $772 \times 10^6 B\bar{B}$ events collected with the Belle detector. The results are
\begin{eqnarray}
	\RD 	&=& 0.307 \pm 0.037 \pm 0.016 \\
	\RDSt   &=& 0.283 \pm 0.018 \pm 0.014,
\end{eqnarray}
where the first uncertainties are statistical and the second are systematic. These results are in agreement with the SM predictions within $0.2\sigma$ and $1.1\sigma$, respectively. The combined result agrees with the SM predictions within $0.8\sigma$. This work constitutes the most precise measurements of \RD\ and \RDSt\ performed to date and the first result for \RD\ based on a semileptonic tagging method. The results of this analysis, together with the most recent Belle results on \RD\ and \RDSt\ (\cite{huschle2015, Hirose:2016wfn}) obtained using a hadronic tag, are combined to provide the Belle combination, which yields $\R(D) = 0.326 \pm 0.034,\ \R(D^*)= 0.283 \pm 0.018$ with a correlation equal to $-0.47$ between the \RD\ and \RDSt\ values. This combined result is in agreement with the SM predictions within 1.6 standard deviations.

We thank the KEKB group for excellent operation of the
accelerator; the KEK cryogenics group for efficient solenoid
operations; and the KEK computer group, the NII, and 
PNNL/EMSL for valuable computing and SINET5 network support.  
We acknowledge support from MEXT, JSPS and Nagoya's TLPRC (Japan);
ARC (Australia); FWF (Austria); NSFC and CCEPP (China); 
MSMT (Czechia); CZF, DFG, EXC153, and VS (Germany);
DST (India); INFN (Italy); 
MOE, MSIP, NRF, RSRI, FLRFAS project, GSDC of KISTI and KREONET/GLORIAD (Korea);
MNiSW and NCN (Poland); MSHE, Agreement 14.W03.31.0026 (Russia); ARRS (Slovenia);
IKERBASQUE (Spain); 
SNSF (Switzerland); MOE and MOST (Taiwan); and DOE and NSF (USA).
We acknowledge the support provided by the Albert Shimmins Fund for the writing of this Letter.


\bibliography{bibliography}

\begin{thebibliography}{30}
\bibitem [{not()}]{note}%
  \BibitemOpen
  \href@noop {} {}\bibinfo {note} {{{Throughout this paper, the inclusion of
  the charge-conjugate decay mode is implied}}}\BibitemShut {NoStop}%
\end{thebibliography}

\end{document}